\documentclass[a4paper,11pt]{article}
\pdfoutput=1 

\usepackage[T1]{fontenc} 
\relax

\usepackage{color}
\usepackage{colordvi}
\usepackage[normalem]{ulem}
\usepackage{color}

\usepackage{graphicx}
\usepackage{epsfig}
\usepackage{graphicx}
\usepackage{longtable}

\usepackage{jheppub} 

\usepackage{hyperref}

\usepackage{amsmath,amssymb,bm}
\usepackage{latexsym}

\newcommand{\mathd}{\mathrm{d}}
\newcommand{\mathi}{\mathrm{i}}

\newcommand\Tr{{\rm Tr }}

\def\Comment#1{}
\newcommand{\bean}{\begin{eqnarray*}}
\newcommand{\eean}{\end{eqnarray*}}

\newcommand{\gapproxeq}{\lower
.7ex\hbox{$\;\stackrel{\textstyle >}{\sim}\;$}}
\newcommand{\lapproxeq}{\lower
.7ex\hbox{$\;\stackrel{\textstyle <}{\sim}\;$}}

\newcommand\lsim{\mathrel{\rlap{\lower4pt\hbox{\hskip1pt$\sim$}}
    \raise1pt\hbox{$<$}}}
\newcommand\gsim{\mathrel{\rlap{\lower4pt\hbox{\hskip1pt$\sim$}}
    \raise1pt\hbox{$>$}}}
\newcommand{\ba}{\begin{array}}
\newcommand{\ea}{\end{array}}
\newcommand{\nn}{\nonumber}

\newcommand{\be}{\begin{equation}}
\newcommand{\ee}{\end{equation}}
\newcommand{\bear}{\begin{eqnarray}}
\newcommand{\eear}{\end{eqnarray}}

\newcommand{\ket}{\,\rangle}
\newcommand{\bra}{\langle \,}
\newcommand{\cO}{{\cal O}}

\newcommand{\mL}{\mathcal{L}}
\newcommand{\mA}{\mathcal{A}}

\newcommand{\mF}{\mathcal{F}}

\newcommand{\mH}{\mathcal{H}}

\def\bat{\begin{array}{cc}}

\newcommand{\Frac}[2]{\frac{\displaystyle #1}{\displaystyle #2}}
\newcommand{\Int}{\displaystyle{\int}}

\preprint{\hfill BARI-TH/2013-673
\\ \phantom{ } \hfill FTUAM-13-4
\\ \phantom{ } \hfill IFT-UAM/CSIC-13-108}

\title{\boldmath Low-energy photon and pion scattering in holographic QCD}

\author[1]{Pietro Colangelo,}
\author[2]{Juan Jose Sanz-Cillero,}
\author[1]{Fen Zuo}
\affiliation[1]{Istituto Nazionale di Fisica Nucleare, Sezione di Bari, Italy}
\affiliation[2]{Departamento de F\'\i sica Te\'orica,  Universidad Aut\'onoma de Madrid, Cantoblanco, 28049 Madrid, Spain}
\emailAdd{Pietro.Colangelo@ba.infn.it}
\emailAdd{juanj.sanz@uam.es}
\emailAdd{Fen.Zuo@ba.infn.it}

\abstract{
Using holographic models where chiral symmetry is broken through IR b.c.'s,  we determine a novel set of relations
between QCD matrix elements. In particular, we find that the amplitudes of the three processes  $\pi\pi\to\pi\pi$, $\gamma\gamma\to\pi\pi$ and $\gamma\to\pi\pi\pi$
involve a single scalar function $h(Q^2)$  given by a suitable 5D integral of
the EoM Green's function.
In a   phenomenological analysis of   $\gamma\gamma\to\pi\pi$ we
 find an overall agreement with the experimental cross section for a broad range of energy. Moreover,
the   polarizabilities at low energies show
a fair agreement between the holographic approach,  previous computations and experiment.
}

\begin{document}
\maketitle
\flushbottom
\section{Introduction}

Inspired by a holographic analysis of the axial-vector-vector ($AVV$) and left-right ($LR$) quark current  Green's functions~\cite{Son:2010vc},
several investigations have been devoted  to the possible interplay between the anomalous and
even intrinsic-parity sectors of Quantum Chromodynamics~\cite{Colangelo:2011xk,Iatrakis:2011ht,Cappiello:2010tu,Domokos:2011dn,Alvares:2011wb,Gorsky:2012ui,Knecht:2011wh}.
In a recent study~\cite{Colangelo:2012ip},   using holographic models where chiral symmetry is realized nonlinearly  through boundary conditions (b.c.'s)~\cite{Son:2003et,Sakai:2004cn,Hirn:2005nr,Sakai:2005yt},
we derived a series of novel form factor  and low-energy constant (LEC) relations
in the limit of large number of colors $N_C\to\infty$~\cite{Nc}.
Here we continue along that line, and extend the analysis to
scattering processes,  going beyond the realm of static properties (mass spectra and couplings) and  facing the more
difficult dynamical problem of   two-body  scattering amplitudes.
We focus   on  $\pi\pi$--scattering
and on the radiative processes  $\gamma\gamma\to \pi\pi$ and $\gamma\to3 \pi$,
finding that it is possible to describe the three amplitudes through a
single function  determined by an appropriate 5D integral
of the Green's function of the five-dimensional equations of motion (EoM).

Chiral Perturbation theory ($\chi$PT) is the effective field theory  describing the low-energy interaction
of the pseudo-Goldstone bosons which emerge from the spontaneous
breaking of the chiral symmetry~\cite{Weinberg-chpt,op4-chpt}.
The observables are obtained by a perturbative expansion in terms of the external momenta and the pseudoscalar mass, and involve a set of effective couplings \cite{Bijnens:1999hw,Bijnens:1999sh,Bijnens:2001bb,Ebertshauser:2001nj}.
However, these  couplings of the low-energy theory are not fixed by the symmetry, but  need to be determined through other
procedures. In~\cite{Colangelo:2012ip} we considered a set of holographic models without explicit chiral
symmetry breaking, and computed all the LEC's of the $\cO(p^6)$ $\chi$PT Lagrangian in the absence of
scalar-pseudoscalar sources. As the LEC's are independent of the quark masses, our results for the
chiral couplings remain valid in the massive quark case.

Here, using  the $\cO(p^6)$ LEC determinations in~\cite{Colangelo:2012ip}, we carry out
a  phenomenological analysis of the
$\gamma\gamma\to\pi\pi$ reaction,
as this matrix element depends on a  transparent way on the $\cO(p^6)$
LEC's~\cite{Burgi:1996,Gasser:2005,Gasser:2006}.
We also study the low-energy polarizabilities
$(\alpha_1+\beta_1)$ and $(\alpha_2\pm \beta_2)$, defined below,
whose lowest chiral orders  are determined by the pion  Born term and the $\cO(p^4)$ one-loop alone, with
no contribution from $\cO(p^4)$ LEC's~\cite{Bijnens:1988}.
The first tree-level contribution
(in addition to  the pion Born term)
occurs at $\cO(p^6)$. Indeed, the polarizabilities
$(\alpha_1+\beta_1)$ and $(\alpha_2+\beta_2)$ vanish at $\cO(p^2)$ and $\cO(p^4)$,
starting the  first non-vanishing contribution at $\cO(p^6)$.
All this makes these observables an interesting benchmark for the
$\cO(p^6)$ LEC determination in~\cite{Colangelo:2012ip}.
We find that the relevant combinations of chiral couplings are fully determined by the anomalous Chern-Simons
action and are universal for this kind of
holographic
models, since they are independent of the  details
of the 5D background~\cite{Colangelo:2012ip}.
In addition, we observe that the experimental $\gamma\gamma\to\pi\pi$ cross section is reproduced
even in the region away from the threshold.

The article is organized as follows.  In Sec.~\ref{sec:sec2} we  recall the holographic setup used  to calculate the LEC and their relations. In Sec.~\ref{sec:sec3}
we show that a unique function appears in several $\pi \pi$ and $\gamma \gamma$ scattering amplitudes,
so that relations can be worked out among these amplitudes and they can be experimentally  tested. In Sec.~\ref{sec:sec4} we study the polarizabilities in $\gamma \gamma \to \pi \pi$ scattering, and compare the results obtained in our holographic approach to a few experimental and theoretical determinations.
Then, we draw our conclusions.

\section{Holographic setup}\label{sec:sec2}

We restrict ourselves to a class of holographic models where chiral symmetry is realized nonlinearly,  through suitable boundary conditions.
This class of models was proposed in ref.~\cite{Son:2003et}, developed in refs.~\cite{Sakai:2004cn,Hirn:2005nr,Sakai:2005yt},
and further studied  in refs.~\cite{Son:2010vc,Colangelo:2012ip}.
The gauge group is $U(n_f)$, and  the $5D$ action is composed by the Yang--Mills (YM) and Chern--Simons (CS) terms, describing
the intrinsic even-parity and the anomalous QCD sectors, respectively~\cite{Son:2003et,Hirn:2005nr,Sakai:2004cn,Sakai:2005yt}:
\begin{equation}
  S = S_{\rm YM}+S_{\rm CS} \label{eq.5Daction}
\end{equation}
with
\begin{eqnarray}
  \label{eq:YM}
  S_{\rm YM} &=& - \int \,\,  d^5x \,\, \Tr \left[-f^2(z){\cal F}_{z\mu}^2
  + \frac{1}{2g^2(z)}{\cal F}_{\mu\nu}^2 \right], \\
  \label{eq:CS}
  S_{\rm CS} &=& -\kappa \int \,\,  \Tr \left[{\cal AF}^2+\frac{i}{2}{\cal A}^3{\cal F}-\frac{1}{10}{\cal A}^5 \right].
\end{eqnarray}
The fifth coordinate $z$ runs from $-z_0$ to $z_0$, with $0<z_0\le+\infty$. ${\cal A}(x,z)={\cal A}_M dx^M$ is the 5D gauge field and
${\cal F}=d{\cal A}-i{\cal A} \wedge {\cal A}$  the field strength. In terms of the $U(n_f)$ generators $t^a$,  normalized as $Tr [t^a t^b]=\delta^{ab}/2$, they read:
 ${\cal A}= {\cal A}^a t^a$ and ${\cal F}= {\cal F}^a t^a$.
The coefficient $\kappa=N_C/(24\pi^2)$ is fixed by the chiral anomaly of
QCD~\cite{Wess:1971yu,Witten:1983tw}.

If the functions $f^2(z)$ and $g^2(z)$ in (\ref{eq:YM}) are invariant under reflection $z\to -z$,
parity can be properly defined in these models.
In appendix~\ref{app.constraints}  other conditions necessary to derive physical  LEC's are discussed.
Moreover, in appendix~\ref{app.holographic-models} we provide the profiles of $f(z)$ and $g(z)$
for the flat metric~\cite{Son:2003et}, ``Cosh''~\cite{Son:2003et},
Hard-Wall~\cite{Hirn:2005nr} and  Sakai-Sugimoto models~\cite{Sakai:2004cn,Sakai:2005yt} that are  used below.
In all these models a common coupling constant $g_5^{-2}$ appears in $f^2(z)$ and $g^{-2}(z)$.
In asymptotically anti-de Sitter backgrounds this coupling can be fixed to be $g_5^{-2}=N_C/(24\pi^2)$~\cite{Son:2003et} through the high-energy behaviour of the two-point correlation function of quark vector currents.
For general backgrounds, both $f^2(z)$ and $g^{-2}(z)$ should always contain a factor of $N_C$,
in order to match the $N_C$-dependence of  the LEC's. For example, one has
\begin{eqnarray}
F^2&=&4\left(\int_{-z_0}^{z_0}\frac{\mathd z}{f^2(z)}\right)^{-1} \, ,
\nn \\
L_1&=&\frac{1}{2} L_2=-\frac{1}{6}L_3=\frac{1}{32}\int_{-z_0}^{z_0} \frac{(1-\psi_0^2)^2}{g^2(z)}  \, \mathd z \, ,
\end{eqnarray}
and all of them are $\cO(N_C)$,  as expected.
%

As discussed in details in ref.~\cite{Colangelo:2012ip}, in (\ref{eq.5Daction}) the QCD chiral symmetry
is promoted to a 5D gauge symmetry, with the possibility of naturally introducing
the right- and left-handed current sources, $r_\mu(x)$ and $\ell_\mu(x)$ respectively.
The Goldstone bosons are contained in the ${\cal A}_z$ component of the gauge field,
and are described through the chiral field $U$,  given by  the  Wilson line
\begin{equation}
U(x^\mu)=\mbox{P} \exp\left\{i\int^{+z_0}_{-z_0} {\cal A}_z(x^\mu,z') dz'\right\}.
\end{equation}
This field  transforms as
\begin{equation}
U(x)\to g_R(x) U(x) g_L^\dagger(x)
\end{equation}
with $g_L(x)$ and $g_R(x)$ the left and right  gauge transformations localized at $z=-z_0$
and $z=z_0$, respectively.   The ${\cal A}_\mu$ components of the gauge fields contain
vector and axial-vector resonances, since
\begin{equation}
{\cal A}_\mu(x,z)=\ell_\mu(x) \psi_-(z)+r_\mu(x) \psi_+(z)
+\sum_{n=1}^\infty B_\mu^{(n)}(x) \psi_n(z)\, .
\label{eq.Amu-decomposition1}
\end{equation}
The ultraviolet (UV) boundary conditions
\be
\mA_\mu(x, -z_0)=\ell_\mu(x)\, ,\qquad\qquad
\mA_\mu(x,z_0)=r_\mu(x)\,
\label{eq.Amu-bcs}
\ee
are imposed. In Eq.(\ref{eq.Amu-decomposition1}) the functions
$\psi_\pm(z)=\frac{1}{2}(1\pm\psi_0(z))$ are determined by  the
$\psi_0(z)$  solution of the 5D EoM  of the $ {\cal A}_\mu(x,z)$ gauge fields
corresponding to the zero mode,
with   b.c.'s   $\psi_0(\pm z_0)= \pm 1$. On the other hand,
the $\psi_n(z)$ correspond to the resonant modes
with mass $m_n$; their  b.c.'s are $\psi_n(\pm z_0)=0$.
Under a suitable gauge transformation it is possible to set $\mA_z=0$, with the space-time  components
of the  5D field taking  the form
\begin{equation}
{\cal A}_\mu(x,z)=i\Gamma_\mu(x)+\frac{u_\mu(x)}{2}\psi_0(z)
+\sum_{n=1}^\infty v_\mu^n(x)\psi_{2n-1}(z)
+\sum_{n=1}^\infty a_\mu^n(x)\psi_{2n}(z)\, ,
\label{eq.Amu-decomposition2}
\end{equation}
where the tensors  $u_\mu(x)$ and $\Gamma_\mu(x)$,
 commonly used  in
$\chi$PT~\cite{Bijnens:1999sh,Bijnens:1999hw,RChTa+b,RChTc},    
    naturally show up:
\begin{eqnarray}
  u_{\mu} \left( x \right)  & \equiv &
  \mathi \left\{ \xi_R^{\dag}\left( x \right)  \left( \partial_{\mu} - \mathi r_{\mu}
  \right) \xi_R\left( x \right)
  - \xi_L^{\dag}\left( x \right)  \left( \partial_{\mu} - \mathi \ell_{\mu}
  \right) \xi_L\left( x \right) \right\} \\
    \Gamma_{\mu} \left( x \right) & \equiv & \frac{1}{2}  \left\{ \xi_R^{\dag}\left( x \right)
  \left( \partial_{\mu} - \mathi r_{\mu} \right) \xi_R\left( x \right) + \xi^{\dag}_L\left( x \right)  \left(
  \partial_{\mu} - \mathi \ell_{\mu} \right) \xi_L\left( x \right) \right\},
\end{eqnarray}
with the non-linear realization $u(x)= \xi_R(x)=\xi_L^\dagger(x)=\exp\{ i \pi^a t^a/F\}$.

The 5D action can be expressed using the decomposition~(\ref{eq.Amu-decomposition2})
of the 5D gauge fields in resonances and the non-linearly realized chiral
Goldstone bosons~\cite{Sakai:2004cn,Hirn:2005nr,Sakai:2005yt,Colangelo:2012ip}.
The resulting 4D action terms relevant for our analysis contain operators
with only Goldstones (see appendix~\ref{app.constraints}) and pieces with
one resonance field $v^n_\mu$,  together with the couplings  $b_{v^n \pi\pi}$, $c_{v^n}$
and $d_{v^n}$:
\begin{eqnarray}
S_{\rm{YM}}\Big|_{1-res}&=& -\,  \sum_n  \frac{i \, b_{v^n\pi\pi}}{4}\,   \int \mathd ^4x
    \bra (\nabla_\mu v_\nu^n-\nabla_\nu v_\mu^n)     [u^\mu,u^\nu]    \ket \,\,\, +\,\,\, ...
\label{eq.S-1res-even}\\
S_{\rm{CS}}\Big|_{1-res} &=&
\, \sum_n  \, \epsilon^{\mu\nu\alpha\beta} \, \int \mathd ^4x \,
\bigg[ -\frac{N_C}{32\pi^2} c_{v^n}
\bra u_\mu\{v_\nu^n,f_{+\alpha\beta}\}\ket +\frac{i N_C}{16\pi^2}\, (c_{v^n}-d_{v^n})
\bra v_\mu^n u_\nu u_\alpha u_\beta\ket \,\bigg]\nn\\
&& +\,\,\, ...
\label{eq.S-1res-odd}
\end{eqnarray}
with $v_\mu^n = v_\mu^{n,\, a} \, t^a$ and the covariant tensor
$f_{+}^{\alpha\beta}=\xi_R^\dagger F_R^{\alpha\beta} \xi_R +
\xi_L^\dagger F_L^{\alpha\beta} \xi_L$ provided by the left and right
source
field-strength tensors~\cite{Bijnens:1999sh,Bijnens:1999hw,RChTa+b,RChTc}.         
The couplings are given by the integrals of the corresponding 5D wave functions:
\begin{eqnarray}
b_{v^n\pi\pi}&=&\int_{-z_0}^{z_0} \frac{\psi_{2n-1}(z)\, (1-\psi_0(z)^2)}{g^2(z)}\, \mathd z \, ,
\nn\\
c_{v^n}&=&-\frac{1}{2}\int _{-z_0}^{z_0}\psi_0(z)\,\psi_{2n-1}'(z) \mathd z\, , \\
d_{v^n}&=&\frac{1}{2}\int_{-z_0}^{z_0}\psi_0(z)^2\, \psi_0'(z)\, \psi_{2n-1}(z)\, \mathd z \, .   \nn
\end{eqnarray}
The primes denote derivative with respect to $z$.
Using the EoM of $\psi_n(z)$ the relation
\begin{eqnarray}
c_{v^n}&=& g_{v^n \pi\pi} \,\,\,
\label{eq.b-c-rel}
\end{eqnarray}
is obtained~\cite{Sakai:2005yt,Colangelo:2012ip},
with $\displaystyle g_{v^n \pi\pi}= \frac{m_{v^n}^2}{2F^2 }b_{v^n\pi\pi}$ the coupling between the  vector resonance $v^n$ and a pion pair.
This equation  implies a connection between the $v^n\to\pi\pi$ and $v^n\to \pi \gamma$ decay widths which are determined by the couplings
$g_{v^n \pi\pi}$ and $c_{v^n}$, respectively, and represents the key ingredient for the relations between amplitudes exploited in the following. Another important property in the considered models is that the vertex $a^n \pi\gamma$ vanishes~\cite{Hirn:2005nr,Colangelo:2012ip}, therefore  all the amplitudes studied  below do not get contributions from the axial-vector resonances.

One may wonder whether the relation (\ref{eq.b-c-rel}) is experimentally fulfilled.
The experimental results for the lightest vector mesons are~\cite{Sakai:2005yt}
\be
c_\omega|_{\rm exp} \,\,\,=\,\,\,  5.80\,\,\,  , \qquad \qquad
g_{\rho\pi\pi}|_{\rm exp} \,\,\,=\,\,\,  5.99\,\,\,\,  ,
\ee
determined  from the  measured  $\omega\to \pi^0\gamma$ and $\rho\to \pi\pi$ decay
rates, respectively~\cite{Beringer:1900zz}.
On the other hand, in the holographic analysis in ref.~\cite{Colangelo:2012ip} we found for
the considered 5D models the  results in Table~\ref{tab.holography-crho},
all close to the experimental data. In each case and all along the paper, the mass
of the lightest vector meson $m_\rho=776$~MeV and the pion decay constant $F=87$~MeV are taken
as inputs to set the parameters of the 5D model~\cite{Colangelo:2012ip}.
Since the  action~(\ref{eq.5Daction}) does not incorporate quark masses,
all the interaction vertices are predicted in the chiral limit.
This allows us to extract some relevant combinations of LEC's, as they are quark mass
independent. However, in the computation of cross sections and decay widths
we  consider the physical pion mass in the phase-space factors,
assuming that this captures the most important quark mass corrections.
Likewise, no scalar states or resonances with spin $S\geq 2$ are included in the present analysis.

\begin{table}[!h]
\centering
\begin{tabular}{|c|c|c|c|c|c|c|}
\hline
&Flat   &  Cosh &  Hard-Wall  &  Sakai-Sugimoto
 \\ \hline
$g_{\rho\pi\pi}=c_\rho$      &  5.11 & 5.14 & 5.13 & 5.11
\\   \hline
\end{tabular}
\caption{{\small
Couplings $g_{\rho\pi\pi}$  and $c_\rho$  determined in four holographic QCD models.
}} 
 \label{tab.holography-crho}
 \end{table}

\section{Holographic description of scattering amplitudes}\label{sec:sec3}
\subsection{$\pi\pi$ scattering}

We consider the $\pi\pi$ scattering in the holographic framework. The scattering amplitude can be easily derived with the 4D resonance Lagrangian (\ref{eq.S-1res-even}), together with the $\cO(p^4)$ chiral Lagrangian resulting from the YM action (\ref{eq:YM}). As we shall see below, the summation over the resonances can be  transformed into a single function $h(Q^2)$ given by an integral involving the 5D Green's function.

The $\pi^a\pi^b\to\pi^c\pi^d$ scattering  amplitude is provided by the isospin decomposition
\bear
A(\pi^a\pi^b \to\pi^c\pi^d) &=& A(s,t,u) \delta^{ab}\delta^{cd} \, +\,
A(t,s,u) \delta^{ac}\delta^{bd} \, +\,
A(u,t,s) \delta^{ad}\delta^{bc} \, ,
\eear
with $s=(p_a+p_b)^2$, $t=(p_a-p_c)^2$, $u=(p_a-p_d)^2$ and the Mandelstam relation
$s+t+u= 0$ in the massless pion limit.
The  holographic action yields the $v^n\pi\pi$ Lagrangian (\ref{eq.S-1res-even}),
in addition to direct $\pi\pi\to \pi\pi$ vertices, giving the amplitude
\bear
A(s,t,u) &=&    \Frac{s}{F^2}
+\Frac{ 4 L_1 ( (t-u)^2 - 3 s^2) }{F^4}
+\sum_n    \Frac{b_{v^n \pi\pi}^2 }{4 F^4} \bigg[ \Frac{t^2 (t+2 s)}{m_{v^n}^2-t}
+\Frac{u^2 ( u+ 2s)}{m_{v^n}^2-u}\bigg]
\nn\\
&=&
\sum_n g_{v^n \pi\pi}^2  \bigg[ \Frac{t+2 s}{m_{v^n}^2-t}
+\Frac{ u+ 2s}{m_{v^n}^2-u}\bigg]\,
\label{eq.Apipi-4D}
\eear
obtained using the definition of
$g_{v^n \pi\pi}$
together with  the
sum rules~\cite{Colangelo:2012ip,Hirn:2005nr}
\be\sum_n b_{v^n\pi\pi}^2 = 32 L_1 \,\,\,\,, \,\,\,\,\,
\sum_n b_{v^n\pi\pi}^2 m_{v^n}^2 =\frac{4 F^2}{3}\, .
\label{eq:sumr}
\ee
It is possible to express the $\pi\pi$--scattering amplitude~(\ref{eq.Apipi-4D}) in a
holographic form,
\bear
A(s,t,u) &=& \Frac{1}{4}\Int_{-z_0}^{z_0}dz\,\, \Int_{-z_0}^{z_0}dz'\,\,
\psi_0'(z)\psi_0'(z')  \,
\bigg[ (t+2s) \, G(-t;z,z') \, + \,  (u+2 s) \, G(-u;z,z') \bigg]
\nn\\
&=& \Frac{1}{4}\,\, \bigg[ (t+2s) \, h(-t) \, + \,  (u+2 s) \, h(-u) \bigg]\, ,
\label{eq.Apipi-5D}
\eear
in terms of the function  $h(Q^2)$ obtained from the integral of the Green's function $G(Q^2;z,z')$ in the $z$ (holographic)  coordinate:
\begin{eqnarray}
h(Q^2) &=& \Int_{-z_0}^{z_0}dz\,\, \Int_{-z_0}^{z_0}dz' \,
\psi_0'(z)\psi_0'(z')  \, G(Q^2;z,z')\, ,
\label{eq.h-def}
\\
G(Q^2;z,z')&=&
\sum_{n=1}^\infty \Frac{\psi_n(z)\psi_n(z')}{m_{n}^2+Q^2} \,.
\label{eq.G-def}
\end{eqnarray}
To rewrite~(\ref{eq.Apipi-4D}) into~(\ref{eq.Apipi-5D}) we made use of the relation
in Eq.~(\ref{eq.b-c-rel}).
It is useful to express the $h(Q^2)$ function in terms of resonance exchanges,
\bear
h(Q^2)  &=&
\sum_n \Frac{4 g_{v^n\pi\pi}^2}{m_{v^n}^2+Q^2}
\,\,\, =\,\,\, \sum_n\Frac{4 c_{v^n}^2}{m_{v^n}^2+Q^2}\, .\label{eq.h-res}
\eear
With the help  of the sum-rules~(\ref{eq:sumr})
 the low-energy expansion can be worked out:
\bear
h(Q^2) &=& \Frac{4}{3F^2} \, - \,  \Frac{ 32 L_1 Q^2}{F^4}
\, +\, \Frac{32 (C_1 + 4 C_3) Q^4}{F^4}
\, +\, \cO(Q^6)\, .
\label{eq.h-chiral-exp}
\eear
We also used the $\cO(p^6)$ relations
 $(C_1+4 C_3)=(3C_3+C_4)=\sum_n F^4 g_{v^n\pi\pi}^2/(8 m_{v^n}^6)$
derived in~\cite{Colangelo:2012ip}. The right low-energy expansion of
the $\pi\pi$--scattering amplitude at large $N_C$~\cite{op4-chpt,op6-pipi-scat} is recovered,
\begin{eqnarray}
A(s,t,u)^{\rm \chi PT} &=& \Frac{s}{F^2}\, +\, \Frac{8 L_1\, (t^2+u^2 -2 s^2)}{F^4}
\, - \, \Frac{8 (C_1 + 4 C_3) (t+u) (t^2+ tu +u^2) }{F^4}
\,\,\, +\,\,\,\cO(p^8)\, .
\nn\\
\end{eqnarray}
Notice that, in the large-$N_C$ limit, in the absence of other resonances,  $L_2$  and $L_3$
are both related to the low-energy constant $L_1$~\cite{RChTa+b}:
$\displaystyle {L_2}/{2}=-{L_3}/{6}=L_1$.
In a similar way, at large--$N_C$  one  has $C_2=0$ and $(C_1+ 4 C_3 )=(3 C_3 + C_4)$
when just vector resonance exchanges are taken into account~\cite{RChTc,Guo:2007}.
This result provides a consistency check of the holographic derivation and of the  determination in~\cite{Colangelo:2012ip} of the $\cO(p^6)$ low energy constants.

The large-$Q^2$ behavior of the function $h(Q^2)$ also determines the high energy
behavior of the scattering amplitudes. This can be obtained from the
resonance expression (\ref{eq.h-res}), and  depends on the convergence of the sum
\begin{equation}
\mH\equiv \sum_{n=0}^\infty ~c_{v^n}^2 \,\,\, .
\end{equation}
As long as this sum converges, one has
\begin{equation}
h(Q^2)\quad \stackrel{Q^2\to\infty}{\longrightarrow}\quad \frac{4\mH}{Q^2} \,\,\, .
\label{eq.h-asy}
\end{equation}
Using the holographic expression for $c_{v^n}$, we can rewrite the sum as a 5D integral,
\begin{equation}
\mH=\frac{1}{4}\int_{-z_0}^{z_0}~[\psi'_0(z)]^2~g^2(z) \mathd z=\frac{F^4}{16}\int_{-z_0}^{z_0}~\frac{g^2(z)}{f^4(z)} \mathd z  \,\,\, .\label{eq.H}
\end{equation}
For arbitrary background functions $f^2(z)$ and $g^2(z)$ this integral may diverge.
However, if we restrict ourselves to the case when the kinetic coefficient $H_1$
of the external sources in the
$\cO(p^4)$ $\chi$PT Lagrangian is
UV--divergent~\footnote{This requirement is not fulfilled in the flat model,
but $\mH$ is still finite in this model.}, we can prove in general that $\mH$ is finite, as shown in
appendix~\ref{app.constraints}.
This is true for all the models listed in appendix~\ref{app.holographic-models},
although the explicit values of the resonance parameters and of
the sum are different in each case. For example, one has
$\mH=g_5^2/2=4m_\rho^2/(\pi^2 F^2)$
and $\mH=g_5^2/3=m_\rho^2/(3 F^2)$ for the Flat and ``Cosh'' models, respectively.
It is interesting to remark that, contrary to other  matrix elements such as the two-point vector current correlation function,
the short-distance power behavior is similar for models with very different backgrounds near the UV boundary. The reason may be that only when we take the momentum square of the external source to be large, as we usually do in the vector correlator
and in form factors involving pions,
we probe the ultraviolet region of the backgrounds.
For instance, in the isospin $I=2$ channel one has
$T^2(s,t,u)=A(t,s,u)+A(u,t,s)$,   and the large-$s$ behaviour of the corresponding  $J=0$ partial wave is
\begin{equation}
T^2_0(s) = -\, \Frac{1}{64\pi s}\Int_{-s}^0\, (t+2 s)\, h(-t)\, d{\rm t}
\quad \stackrel{s\to\infty}{\longrightarrow}\quad
-\frac{\mH}{8\pi}\ln \frac{s}{\Lambda^2}+\cO(s^0)\, .
\label{eq.t20-log}
\end{equation}
The  parameter $\Lambda$ is a finite hadronic scale  arising from  the integration.
Thus, the residual logarithmic behavior found at high energies
in the case of a finite number of vector
exchanges~\cite{Guo:2007,Nieves:2011}
shows up also in holographic models, with  an infinite tower of resonances. Calculations of
higher partial waves give a similar logarithmic behavior.

\subsection{$\gamma\gamma\to \pi\pi$ scattering}

We now  analyze an observable for which  the odd intrinsic-parity sector of the action
plays a crucial role, the radiative process $\gamma\gamma\to\pi\pi$.
We first perform a separate theoretical study of the neutral and charged modes.
Nonetheless, our holographic description provides pretty similar structures for both of them.

\subsubsection{$\gamma\gamma\to\pi^0\pi^0$}

The $\gamma(k_1,\epsilon_1)\gamma(k_2,\epsilon_2)\to\pi^0(p_1)\pi^0(p_2)$
scattering amplitude is described by
two structure functions, $A(s,t,u)^{\gamma\gamma\to \pi^0\pi^0}$ and
$B(s,t,u)^{\gamma\gamma\to \pi^0\pi^0}$:
\bear
T^{\gamma\gamma\to\pi^0\pi^0}
&=&
e^2   (\epsilon_1^\mu\epsilon_2^\nu T^{(1)}_{\mu\nu})
\times A(s,t,u)^{\gamma\gamma\to\pi^0\pi^0} \,\,\, +\,\,\,
e^2   (\epsilon_1^\mu\epsilon_2^\nu T^{(2)}_{\mu\nu})
\times B(s,t,u)^{\gamma\gamma\to\pi^0\pi^0}\, .
 \nn\\ \label{eq:ggamplitude}
\eear
The Lorentz structures  are defined as  $(\epsilon_1^\mu\epsilon_2^\nu T^{(1)}_{\mu\nu}) =
\frac{s}{2} (\epsilon_1 \epsilon_2) - (\epsilon_1 k_2)(\epsilon_2 k_1)$ and
$(\epsilon_1^\mu\epsilon_2^\nu T^{(2)}_{\mu\nu}) =2 s (\epsilon_1 \Delta)(\epsilon_2 \Delta)
- (t-u)^2 (\epsilon_1\epsilon_2)
- 2 (t-u) [(\epsilon_1\Delta)(\epsilon_2 k_1) - (\epsilon_1 k_2) (\epsilon_2 \Delta)]$,
with  $\Delta^\mu\equiv p_1^\mu -p_2^\mu$ and $\epsilon_i$ the photon polarization vectors.

In $\chi$PT, the neutral $\pi \pi $ channel has no contribution at $\cO(p^2)$,
and at $\cO(p^4)$ there is  no tree-level, but only one-loop diagrams.
Thus, at large-$N_C$ the contributions to the $\gamma\gamma\to\pi^0\pi^0$
amplitude start at $\cO(p^6)$ and
read~\cite{Burgi:1996,Gasser:2005}
\bear
A(s,t,u)^{\gamma\gamma\to\pi^0\pi^0} &=& \Frac{1}{(4\pi F)^4}\, (a_1 m_\pi^2 + a_2 s)\,\,\,
+ \,\,\, \cO(E^4) \,  ,
\nn\\
B(s,t,u)^{\gamma\gamma\to\pi^0\pi^0} &=& \Frac{b}{(4\pi F)^4}\,\,\,  + \,\,\, \cO(E^2)\, , \label{eq:AB}
\eear
with $s,t,u,m_\pi^2=\cO(E^2)$.
Since our holographic approach does not incorporate quark masses, the term with $a_1$  is out of the
reach of the present work, and we focus on observables that do not contain it.
In (\ref{eq:AB})  the parameters $a_2$ and $b$
are combinations of LEC's.  For three flavors and symmetry group $SU(3)$,
the electric charge matrix is Q=diag$(2/3,-1/3,-1/3)$,  and one finds the large--$N_C$
low-energy parameters
\bear
a_2 &=& \Frac{10}{9}\times 256\pi^4 F^2 \bigg(
8 C_{53} + 8 C_{55} + C_{56}+C_{57}+ 2 C_{59}  \bigg)\,
\nn\\
b &=&  -\, \Frac{10}{9}\times 128 \pi^4 F^2 \bigg(
C_{56}+C_{57}+ 2 C_{59}  \bigg)\, .
\eear
In the holographic calculation~\cite{Colangelo:2012ip} we obtained  predictions for the $\cO(p^6)$
LEC's, which   yield   particularly simple expressions,
\bear
a_2 &=& \Frac{10}{9}\times  3 N_C^2 F^2 \sum_n \Frac{c_{v^n}^2}{m_{v^n}^2}
\,\,\,=\,\,\, \Frac{10}{9}\times N_C^2\,
\nn\\
b &=& \Frac{10}{9}\times  \Frac{N_C^2 F^2}{2}  \sum_n \Frac{c_{v^n}^2}{m_{v^n}^2}
\,\,\,=\,\,\, \Frac{10}{9}\times \Frac{N_C^2}{6}\,
\label{eq.neutral-a2-b-holography}
\eear
using the sum   rule $\sum_n c_{v^n}^2 /m_{v^n}^2 = \sum_n g_{v^n \pi\pi}^2/m_{v^n}^2
= 1/(3 F^2)$ derived in~\cite{Colangelo:2012ip}. The expressions for the relevant LEC's  are collected in appendix~\ref{app.LECe}.

At higher energy  it is possible to express our holographic result for the massless quark case in terms
of the  same function  $h(Q^2)$:
\bear
A(s,t,u)^{\gamma\gamma\to\pi^0\pi^0}  &=&
\Frac{a_2}{(4\pi F)^4} \bigg[ \Frac{(s-4t)}{2} \, \sum_n \Frac{F^2 c_{v^n}^2}{m_{v^n}^2-t}
\,\,\, +\,\,\, (t\leftrightarrow u)\bigg] \nn\\
&=&  \Frac{a_2}{(4\pi F)^4} \bigg[ \Frac{(s-4t)}{8} \, F^2 h(-t)  \,\,\, +\,\,\, (t\leftrightarrow u)\bigg] \nn\\
&=&  \Frac{a_2}{(4\pi F)^4} \bigg[ s\,\, +\,\, \Frac{4 L_1}{F^2}(8 t u - 5s^2) \,\, +\,\,\cO(E^6)\bigg]\, ,
\\ \nn\\
B(s,t,u)^{\gamma\gamma\to\pi^0\pi^0}  &=&
\Frac{b}{(4\pi F)^4} \bigg[ \Frac{3}{2} \, \sum_n \Frac{F^2 c_{v^n}^2}{m_{v^n}^2-t}
\,\,\, +\,\,\, (t\leftrightarrow u)\bigg] \nn\\
&=& \Frac{b}{(4\pi F)^4} \bigg[ \Frac{3}{8} \, F^2 h(-t)  \,\,\, +\,\,\, (t\leftrightarrow u)\bigg] \nn\\
&=& \Frac{b}{(4\pi F)^4} \bigg[ 1\,\, -\,\, \Frac{12 L_1 s}{F^2} \,\,+ \,\, \cO(E^4) \bigg] \, ,
\label{eq.holo-gg-pipi-neutral}
\eear
with $a_2$ and $b$ given in Eq.~(\ref{eq.neutral-a2-b-holography})
and the low-energy expansion of $h(Q^2)$ in~(\ref{eq.h-chiral-exp}).
We have kept the chiral expansion up to $\cO(p^8)$,
as later we will analyze  the polarizability $(\alpha_2+\beta_2)$
which starts at that order at large--$N_C$.

For $N_C\to \infty$ the strange quark does not play a role in this amplitude
and our holographic description yields the same prediction in $U(2)$
and $U(3)$. Moreover, since in the external legs   we only have pions and non-singlet components in the
electromagnetic gauge field ($J_\mu^{\rm em}=J_\mu^3 +\frac{1}{\sqrt{3}} J_\mu^8$),
the $U(3)$ result is identical to the $SU(3)$ one.
It is clarifying to observe the $\gamma\to \pi V$ vertex  obtained from the Lagrangian~(\ref{eq.S-1res-odd})
in the $SU(3)$ case,
\bear
\mL &=& \sum_n  \Frac{e N_C c_{v^n}}{8\pi^2 F}\,  \epsilon^{\rho\sigma\mu\nu}
\partial_{\rho} V_{\sigma}^{\rm em}  \, \bigg[
\Frac{1}{3} \bigg(\partial_\mu \pi^0 \, \rho^{0,n}_\nu
+   \partial_\mu \pi^+ \, \rho^{-,n}_\nu
+   \partial_\mu \pi^- \, \rho^{+,n}_\nu \bigg)
+    \partial_\mu \pi^0 \, \omega^{n}_\nu \bigg] \,\,\,+\,\,\, ...
\nn\\
\label{eq.gVpi}
\eear
coincides with the one  found in previous  approaches based on resonance
Lagrangians~\cite{Ko:1990,Babusci:1993}.

If we restrict ourselves to $SU(2)$
with Q=diag$(1/2,-1/2)$  (but allowing the $SU(2)$ singlet
resonance $\omega$),  all the $\rho^{\pm,0}$ terms vanish and only the $\omega$
resonance exchange survives. Thus, the $\cO(p^6)$ $\chi$PT contribution
at large $N_C$~\cite{Gasser:2005},
\bear
a_2 &=&   256\pi^4 F^2 \bigg(
8 C_{53} + 8 C_{55} + C_{56}+C_{57}+ 2 C_{59}  \bigg) \, ,
\nn\\
b &=&  -\,  128 \pi^4 F^2 \bigg(
C_{56}+C_{57}+ 2 C_{59}  \bigg)\, ,
\eear
becomes~\cite{Colangelo:2012ip}
\bear
a_2 = N_C^2 \, \,\,\, ,  \qquad \qquad
b = \Frac{N_C^2}{6} \,\,\,\, .
\label{eq.neutral-a2-b-holography2}
\eear
Notice that  we have used the $SU(3)$ notation for the corresponding
$SU(2)$ couplings at large $N_C$. In $SU(2)$ notation one would have to make the replacement
$C_{53}\to c_{29}$, $C_{55}\to c_{30}$, $C_{56}\to c_{31}$,
$C_{57}\to c_{32}$ and  $C_{59}\to c_{33}$~\cite{Bijnens:2001bb}.
In the  following   phenomenological analysis we  always consider the large--$N_C$
estimates with the charge matrix Q=diag$(2/3,-1/3,-1/3)$.

\subsubsection{$\gamma\gamma\to\pi^+\pi^-$}

The charged pion mode contains the Born term at $\cO(p^2)$ in the chiral expansion,
given by the pion exchange diagram.
At $\cO(p^4)$ there is a tree-level contribution proportional to $(L_9+L_{10})$ and one-loop diagrams.
Hence, at large-$N_C$ (where loop diagrams provide a subleading contribution)
we have the chiral
expansion~\cite{Burgi:1996,Bijnens:1988,Gasser:2005,Gasser:2006}:
\bear
A(s,t,u)^{\gamma\gamma\to\pi^+\pi^-} &=&
\Frac{1}{m_\pi^2-t}+\Frac{1}{m_\pi^2-u}
\,\,\, +\,\,\, \Frac{8(L_9+L_{10})}{F^2} \,\,\,+\,\,\, \Frac{(a_1 m_\pi^2+a_2 s)}{(4\pi F)^4}\,\,\,  + \,\,\, \cO(E^4)\, ,
\nn\\
B(s,t,u)^{\gamma\gamma\to\pi^+\pi^-} &=& \Frac{1}{2s}\bigg[\Frac{1}{m_\pi^2-t}+\Frac{1}{m_\pi^2-u}\bigg]\,\,\,
+\,\,\, \Frac{b}{(4\pi F)^4}\,\,\,  + \,\,\, \cO(E^2)\, ,
\eear
with $s,t,u,m_\pi^2=\cO(E^2)$. In the three-flavor $SU(3)$ case,
$a_2$ and $b$ are given by the combinations of $\cO(p^6)$ LEC's,
\bear
a_2 &=& \Frac{10}{9}\times 256\pi^4 F^2 \bigg(
8 C_{53} - \frac{4}{5}\times 8 C_{55} + C_{56}+C_{57}- \frac{4}{5}\times 2 C_{59}
\nn\\
&&\qquad \qquad \qquad\qquad
+\frac{9}{10}\times 4 C_{78} +\frac{9}{10}\times 8 C_{87} -\frac{9}{10}\times 4 C_{88}
\bigg)\, ,
\nn\\
b &=&  -\, \Frac{10}{9}\times 128 \pi^4 F^2 \bigg(
 C_{56}+C_{57}-\frac{4}{5}\times  2 C_{59}  - \frac{9}{10}\times 4 C_{78}\bigg)\, .
\eear
Using the results for the $\cO(p^6)$
LEC's in~\cite{Colangelo:2012ip}~(collected in appendix~\ref{app.LECe}), we obtain the  low-energy predictions
\bear
a_2 &=& \Frac{1}{9}\times  3 N_C^2 F^2 \sum_n \Frac{c_{v^n}^2}{m_{v^n}^2}
\,\,\,=\,\,\, \Frac{1}{9}\times N_C^2\,\,\, ,
\nn\\
b &=& \Frac{1}{9}\times  \Frac{N_C^2 F^2}{2}  \sum_n \Frac{c_{v^n}^2}{m_{v^n}^2}
\,\,\,=\,\,\, \Frac{1}{9}\times \Frac{N_C^2}{6}\,\,\, .
\label{eq.charged-a2-b-holography}
\eear

At higher energies it is possible to express the  holographic result for massless quarks
in terms of the  function $h(Q^2)$:
\bear
\bar{A}(s,t,u)^{\gamma\gamma\to\pi^+\pi^-}
&=&  \Frac{a_2}{(4\pi F)^4} \bigg[ \Frac{(s-4t)}{2} \, \sum_n \Frac{F^2 c_{v^n}^2}{m_{v^n}^2-t}
\,\,\, +\,\,\, (t\leftrightarrow u)\bigg]
\nn\\
&&   =   \Frac{a_2}{(4\pi F)^4} \bigg[ \Frac{(s-4t)}{8} \, F^2 h(-t)
\,\,\, +\,\,\, (t\leftrightarrow u)\bigg]
\nn\\
&&   =
\Frac{a_2}{(4\pi F)^4} \bigg[ s\,\, +\,\, \Frac{4 L_1}{F^2}(8 t u - 5s^2)
\,\,+ \,\, \cO(E^6)\bigg]\, ,
\nn\\
\nn\\
\bar{B}(s,t,u)^{\gamma\gamma\to\pi^+\pi^-}
&=& \Frac{b}{(4\pi F)^4} \bigg[ \Frac{3}{2} \, \sum_n \Frac{F^2 c_{v^n}^2}{m_{v^n}^2-t}
\,\,\, +\,\,\, (t\leftrightarrow u)\bigg]
\nn\\
&&  = \Frac{b}{(4\pi F)^4} \bigg[ \Frac{3}{8} \, F^2 h(-t)
\,\,\, +\,\,\, (t\leftrightarrow u)\bigg]
\nn\\
&&  =
\Frac{b}{(4\pi F)^4} \bigg[ 1\,\, -\,\, \Frac{12 L_1 s}{F^2} \,\,+ \,\, \cO(E^4) \bigg]\, ,
\label{eq.holo-gg-pipi-charged}
\eear
where the $\cO(p^2)$ pion exchange term has been removed in $\bar{A}^{\gamma\gamma\to\pi^+\pi^-}$
and $\bar{B}^{\gamma\gamma\to\pi^+\pi^-}$, the $\cO(p^4)$ result
$L_9+L_{10}=0$ has been used, and $a_2$ and $b$ are given in Eq.~(\ref{eq.charged-a2-b-holography}).

As  in the neutral channel, the strange quark does not play a role in this amplitude
at  large-$N_C$,   and the  holographic description yields the same prediction in $U(2)$,
$SU(3)$ and $U(3)$.    However, if we restrict ourselves to $SU(2)$ sources and take just the
tripet component of the electromagnetic charge matrix
Q=diag$(1/2,-1/2)$  there is no vector exchange, and
one finds~\cite{Colangelo:2012ip,Gasser:2006}
\bear
a_2 &=&    256\pi^4 F^2 \bigg(
8 C_{53} -  8 C_{55} + C_{56}+C_{57}-   2 C_{59}
+ 4 C_{78} +  8 C_{87} - 4 C_{88} \bigg) \,\,\,=\,\,\, 0\, ,
\nn\\
b &=&  -\,  128 \pi^4 F^2 \bigg(
 C_{56}+C_{57}-  2 C_{59}  -   4 C_{78}\bigg)\,\,\,=\,\,\, 0\, .
\label{eq.charged-a2-b-holography2}
\eear
In these two equations  we have used the $SU(3)$ notation for the corresponding
$SU(2)$ couplings; in $SU(2)$ notation one should replace
$C_{53}\to c_{29}$, $C_{55}\to c_{30}$, $C_{56}\to c_{31}$,
$C_{57}\to c_{32}$,  $C_{59}\to c_{33}$, $C_{78}\to c_{44}$, $C_{87}\to c_{50}$
and $C_{88}\to c_{51}$~\cite{Bijnens:2001bb}.

\subsection{$\gamma\to\pi^+\pi^-\pi^0$ amplitude}

The amplitude  $\gamma^*\to\pi^+\pi^-\pi^0$ is described
in the form~\cite{g-3pi-op4,g-3pi-op6,Hoferichter:2012pm}
\begin{eqnarray}
\bra 0 | J_\mu^{EM}|\pi^+(p_1)\pi^-(p_2)\pi^0(p_3)\ket\,\,\,=\,\,\,
i\, \epsilon_{\mu\nu\alpha\beta} p_1^\nu p_2^\alpha p_3^\beta \, \mF^{3\pi}(q^2,s,t)\, ,
\end{eqnarray}
with $q=p_1+p_2+p_3$ and $s=(p_1+p_2)^2$, $t=(p_1+p_3)^2$,
$u=(p_2+p_3)^2$
(with  $s+t+u=q^2$ in the massless pion case).
In the holographic models, the second term in Eq.~(\ref{eq.S-1res-odd}) gives the direct coupling of the vector meson to three pions, which  contributes to $\mF^{3\pi}(q^2,s,t)$ for $q^2\ne0$.
When the photon is on-shell  ($q^2=0$) the amplitude has a simple structure in the chiral limit:
\bear
\mF^{3\pi}(0,s,t) &=& \mF^{3\pi}_0 \, \times\, \Frac{F^2}{4}\, \bigg[\, h(-s)\, + \, h(-t) \,
+\, h(-u)\,  \bigg]\,
\eear
with $\mF_0^{3\pi}= \frac{e N_C}{12\pi^2 F^3}$~\cite{g-3pi-op4,g-3pi-op6,Hoferichter:2012pm}.
At low energies we recover the ChPT expression
\bear
\mF^{3\pi}(0,s,t) &=& \mF^{3\pi}_0 \, \times\, \bigg[\, 1\,\,\, +\,\,\, \cO(E^4)\,  \bigg]\, ,
\eear
since the $\cO(E^2)$ term cancels due to the  relation $s+t+u=0$ for massless pions.

In principle, further analyses could be considered for the decay $\eta\to \gamma \pi^+\pi^-$.
The study of this  flavor structure might allow the extraction of information about the $\cO(p^6)$ couplings $C_{13,14,15}^W$,
relevant for this kind of radiative processes~\cite{g-3pi-op6}.

\subsection{Relations among the scattering amplitudes through holography}

Before proceeding with the phenomenological analysis,
it is interesting to summarize the holographic results for the massless quark limit,
remarking how the various amplitudes are provided by the same
Green's function  integral $h(Q^2)$.

\begin{itemize}

\item[]{\bf $\pi^+\pi^- \to\pi^0\pi^0$ scattering amplitude}

\bear
A(s,t,u) &=& \Frac{(t+2s)}{4}\, h(-t) \,\,\,+\,\,\ (t\leftrightarrow u) \, .
\eear

\item[]{\bf $\gamma\gamma\to \pi^0\pi^0$ amplitude}

\bear
A(s,t,u)^{\gamma\gamma\to\pi^0\pi^0} &=&  \Frac{a_2}{(4\pi F)^4} \bigg[ \Frac{(s-4t)}{8} \, F^2 h(-t)
\,\,\, +\,\,\, (t\leftrightarrow u)\bigg] \, ,
\nn\\
B(s,t,u)^{\gamma\gamma\to\pi^0\pi^0} &=&  \Frac{b}{(4\pi F)^4} \bigg[ \Frac{3}{8} \, F^2 h(-t)
\,\,\, +\,\,\, (t\leftrightarrow u)\bigg]\, .
\eear
The amplitude with charged pions has the same structure, up to
a global factor, with the addition  of the Born pion-exchange term.

\item[]{\bf $\gamma\to\pi^+\pi^-\pi^0$ radiative process}

\bear
\mF^{3\pi}(0,s,t) &=&
\mF^{3\pi}_0 \, \times\, \Frac{F^2}{4}\, \bigg[\, h(-s)\, + \, h(-t) \,
+\, h(-u)\,  \bigg]\, .
\eear
\end{itemize}

All the above scattering amplitudes are calculated in the 4D picture, i.e., using the resonances expansion.
Since the results are all expressed in terms of  the 5D Green's function, it would be interesting to see if they
can be directly  obtained  from the five-dimensional action, a derivation still missing at present.

\section{Polarizabilities and cross sections for  $\gamma\gamma\to \pi\pi$ scattering}\label{sec:sec4}

The polarizabilities can be defined  following  the notations provided by Refs.~\cite{Gasser:2005,Gasser:2006}.
The helicity amplitudes are written as
\bear
H_{++} &=& A(s,t,u)^{\gamma\gamma\to\pi\pi}\, +\,  2 (4 m_\pi^2 -s) B(s,t,u)^{\gamma\gamma\to\pi\pi}\, ,
\nn\\
H_{+-} &=& \Frac{ 8( m_\pi^4 -t u )}{s} B(s,t,u)^{\gamma\gamma\to\pi\pi}\, .
\eear
which determine the  $\gamma\gamma\to\pi^+\pi^-$ cross section~\cite{Burgi:1996,Gasser:2006}
\be
\sigma = \Frac{\alpha^2 \pi}{8} \Int_{t_-}^{t_+} dt\, H(s,t)\, ,     
\label{eq.cross-section}
\ee
with $H= |H_{++}|^2 \,+\, |H_{+-}|^2$,  $t_\pm = m_\pi^2 -\Frac{s}{2} (1\mp \beta(s) Z)$, $\beta(s)=\sqrt{1-4 m_\pi^2/s}$,
for scattering angles in the range $|\cos\theta|<Z\leq 1$. The Mandelstam variables are
$t=(p_1-k_1)^2= m_\pi^2-\Frac{s}{2}(1-\beta(s)\cos\theta)$
and $u=(p_2-k_1)^2= m_\pi^2-\Frac{s}{2}(1+\beta(s)\cos\theta)$,
with $\vec{k}_1 \cdot \vec{p}_1 = |\vec{k}_1|\, |\vec{p}_1|\, \cos\theta$.
They obey the  relation $s+t+u= 2 m_\pi^2$.
The  $\gamma\gamma\to\pi^0\pi^0$ cross section is given by Eq.~\eqref{eq.cross-section} with an 
additional $1/2$ factor~\cite{Gasser:2005}.

\begin{figure}[!t]\centering
\includegraphics[width=12cm,angle=0,clip]{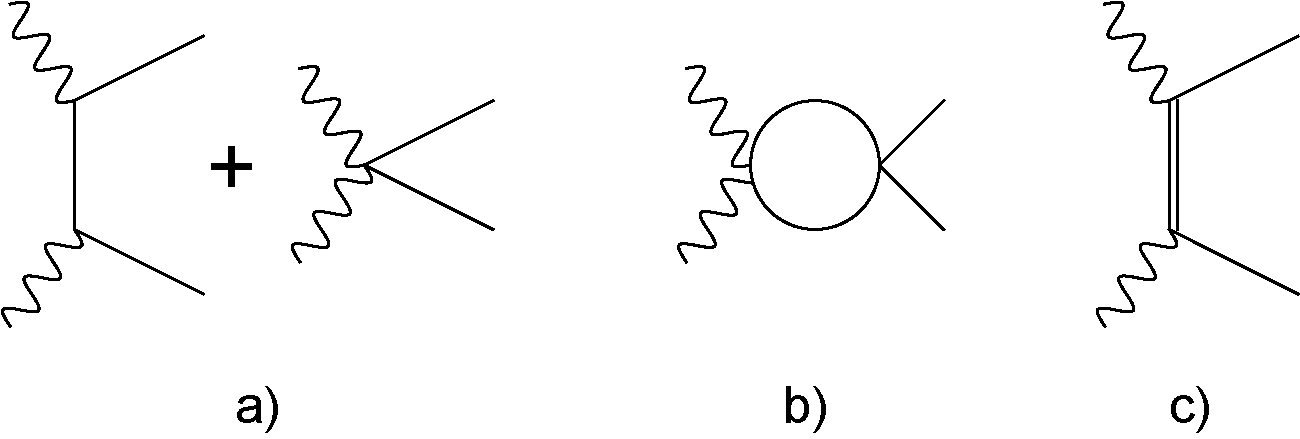}
\caption{\small {
Feynman diagrams for $\gamma\gamma\to\pi\pi$:
a) $\cO(p^2)$  and $\cO(p^4)$ tree-level (only in the charged channel),
b) example of $\cO(p^4)$ loops, c) vector resonance exchanges
($\cO(p^6)$ and higher in our holographic approach). No other resonance diagram contributes in our analysis.
The wavy, solid and double lines stand for photons, pions and vector resonances, respectively.
}}
\label{fig.diagrams}
\end{figure}

\begin{figure}[!t]\centering
\includegraphics[width=10cm,angle=0,clip]{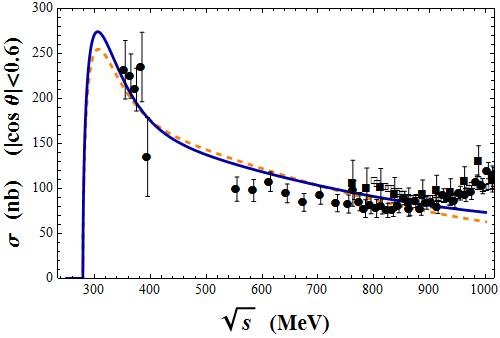}
\caption{\small {
$\gamma\gamma\to\pi^+\pi^-$ cross section
compared to our large-$N_C$ expressions.  The dashed (orange) line is obtained from
the Born term with  one-pion exchange,   which is $\cO(p^2)$ in the
$\chi$PT counting. In the holographic approach we  have also vector exchanges,
but their effect is essentially negligible even at high energies.
The solid (blue) curve represents the prediction obtained including
the Born term + $\cO(p^4)$ loop + vector exchanges.
The  data points are from MARK-II (circles)~\cite{MARK-II}, CELLO (filled squares)~\cite{CELLO},
and BELLE experiment (empty squares)~\cite{charged-BELLE}.
}}
\label{fig.sigma-charged}
\end{figure}

The charged channel  is
given   at $\cO(p^2)$ in $\chi$PT   by the Born term, $A^{\rm \small Born}
= \frac{1}{m_\pi^2-t}+\frac{1}{m_\pi^2-u}= 2 \, s \, B^{\rm Born}$
(Fig.~\ref{fig.diagrams}a).
This  already provides
a fairly good description of the experimental $\gamma\gamma\to\pi^+\pi^-$
data,  as one can see in Fig.~\ref{fig.sigma-charged}.
At $\cO(p^4)$ there is a one-loop contribution
and a tree-level term proportional to $L_9+L_{10}$
(appendix~\ref{app.Op4})~\cite{Bijnens:1988,Gasser:2006}.
The latter happens to be zero in our large $N_C$ holographic
models~\cite{Hirn:2005nr,Sakai:2005yt}.
%
%
Likewise, one also has vector resonance exchanges
in the crossed channel
(Fig.~\ref{fig.diagrams}c),
which start contributing at $\cO(p^6)$ at  low momenta.
However, it is possible to observe in
Fig.~\ref{fig.sigma-charged} that these corrections  to the Born term
are tiny even up to energies of the order of $\sqrt{s}\sim 1$~GeV,
where one starts being sensitive
to the $s$--channel resonances $f_0(980)$ and $f_2(1270)$.
Nonetheless, we notice  that in the charged channel the vector resonance exchanges
(leading in $1/N_C$ but starting
at NNLO in the chiral counting at very low energies) are much smaller
than the $\cO(p^4)$ loop
(subleading in $1/N_C$ but NLO in the chiral expansion):
if the vector exchanges were removed
it would not be possible to see the difference
with the full result (Born+$\cO(p^4)$ loop+vector exchanges)
in Fig.~\ref{fig.sigma-charged}.
However, as we discuss  below, this is no longer true for  the
helicity amplitude $H_{+-}$ and the corresponding low-energy polarizabilities
$(\alpha_1+\beta_1)_{\pi^+}$  and  $(\alpha_2+\beta_2)_{\pi^+}$,
which receive their first non-vanishing contribution
at $\cO(p^6)$ in the chiral expansion.

\begin{figure}[!b]\centering
\includegraphics[width=10cm,angle=0,clip]{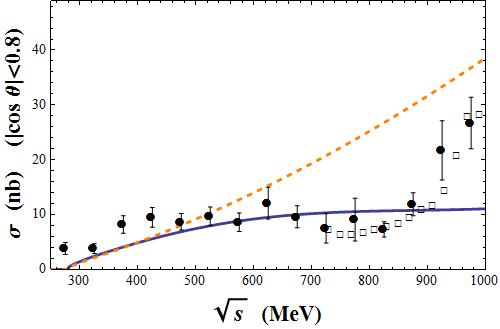}
\caption{\small {
$\gamma\gamma\to\pi^0\pi^0$ cross section.
Data from Crystal Ball~\cite{CrystalBall:1990} (circles) and BELLE~\cite{neutral-BELLE} (empty squares)
are compared to the $\cO(p^4)$
$\chi$PT expression, which comes only  from one-loop diagrams (dashed orange).
If we add  the resonance contributions from the holographic ``Cosh'' model,
we obtain the solid blue curve.  One can appreciate the relevance of the pion loops at low energies,
together with the importance of its interference with the vector meson exchanges
at higher energy.
}}
\label{fig.sigma-neutral}
\end{figure}

At $\cO(p^2)$ in $\chi$PT  the neutral channel amplitude is zero.
It gets its first contribution at $\cO(p^4)$
in the chiral expansion
via one-loop diagrams (Fig.~\ref{fig.diagrams}b).
Its expression is provided in appendix~\ref{app.Op4}~\cite{Bijnens:1988,Gasser:2005}.
No tree-level diagram contributes at this order, and the loops
are UV finite.  Hence, there is  a competition between
the dominant chiral order in $\gamma\gamma\to\pi^0\pi^0$, $\cO(p^4)$,
which is subleading in $1/N_C$, and the  dominant contribution
at large--$N_C$, given by the vector exchanges and starting at $\cO(p^6)$.
As one can see in Fig.~\ref{fig.sigma-neutral}, near threshold the
one-loop $\cO(p^4)$~\cite{Bijnens:1988} contribution dominates, but as the energy increases the
tree and loop diagrams interfere and the description  improves at higher
energies. We remark the importance of this interference, since just the resonance
exchanges undervalue the cross section in the energy range below 800~MeV.

Data from MARK-II~\cite{MARK-II}, CELLO~\cite{CELLO}, Crystal Ball~\cite{CrystalBall:1990} and
BELLE~\cite{charged-BELLE,neutral-BELLE} Collaborations are compared to our theoretical estimates in Figs.~\ref{fig.sigma-charged}~and~\ref{fig.sigma-neutral}.
In particular, we provide the holographic results for the ``Cosh'' model; the  results for the Hard-Wall model are practically identical
and could not be distinguished in the plots. In the charged channel case, where
the loop and crossed vector exchanges are found to be very tiny, the first significant
difference is expected around $\sqrt{s}\sim 1$~GeV where the effect of the
$s$--channel resonance $f_0(980)$ and
$f_2(1270)$ starts being relevant. It is interesting to observe that
although the scalar $f_0(600)$ may explain the small
deviations from the experimental data below $1$ GeV,
it  does not play a significant role. On the other hand,
the large crossed pion exchange contribution is absent
in the $\gamma\gamma\to\pi^0\pi^0$  cross section,   and
the pion loops and their interference with the crossed vector contributions
are crucial. A more precise analysis for $\sqrt{s}<1$~GeV would require
a detailed study of  $\pi\pi$ final state interactions.

The low-energy expansion of the helicity amplitudes for $t=m_\pi^2$ provides the
polarizabilities~\cite{Gasser:2005,Gasser:2006}:
\bear
\Frac{\alpha}{m_\pi} H_{++ }(s,t=m_\pi^2) &=&
\Frac{\alpha}{m_\pi}\, \bigg[A^{\gamma\gamma\to\pi\pi} \, \, +\,\,  2 (4 m_\pi^2 -s) B^{\gamma\gamma\to\pi\pi}\,
\bigg]_{t=m_\pi^2}
\nn\\
& \,\,\,\stackrel{s\to 0}{=} \,\,\,  &
(\alpha_1- \beta_1)\,\,\, +\,\,\, \Frac{s}{12}\, (\alpha_2- \beta_2)\, \,\, +\,\,\,
\cO(s^2)\, ,
\nn\\
\nn\\
\Frac{\alpha}{m_\pi} H_{+- }(s,t=m_\pi^2) &=&
\Frac{\alpha}{m_\pi}\, 8 m_\pi^2  B^{\gamma\gamma\to\pi\pi}\bigg|_{t=m_\pi^2}
\nn\\
&\,\,\,\stackrel{s\to 0}{=} \,\,\, &
(\alpha_1+ \beta_1)\,\,\, +\,\,\, \Frac{s}{12}\, (\alpha_2+\beta_2)\, \,\, +\,\,\,
\cO(s^2)\, .
\eear
In the case of the charged pion amplitude,  the Born term is subtracted  when
defining the polarizabilities.

The  chiral expansion for the $\gamma\gamma\to\pi^0\pi^0$ polarizabilities
are~\cite{Gasser:2005}:
\bear
(\alpha_1-\beta_1)_{\pi^0} &=& \Frac{\alpha}{m_\pi} \,\Frac{1}{(4\pi F)^2}\,
\bigg\{  \, -\, \Frac{1}{3}\,\, +\,\, \, \left(a_1^r+ 8 b^r + \cO(p^6) \mbox{ loops}\right) \,
 \Frac{m_\pi^2}{(4\pi F)^2} \, \, + \,\,\cO(m_\pi^4)\bigg\}\, ,
\nn\\
(\alpha_1+\beta_1)_{\pi^0} &=& \Frac{\alpha}{m_\pi}\, \Frac{m_\pi^2}{(4\pi F)^4}  \,
\bigg\{ \, \left(8b^r +  \cO(p^6) \mbox{ loops} \right)
\,  \, + \,\, \cO(m_\pi^2)\bigg\}\, ,
\nn\\
(\alpha_2-\beta_2)_{\pi^0} &=& \Frac{\alpha}{m_\pi}\, \Frac{1}{m_\pi^2 (4\pi F)^2} \,
\bigg\{ \Frac{156}{45}\, +\,
\left(12 a_2^r -24 b^r +  \cO(p^6) \mbox{ loops} \right)\, \Frac{m_\pi^2}{(4\pi F)^2}  \, +
\,\cO(m_\pi^4)\bigg\}\, ,
\nn\\
(\alpha_2+\beta_2)_{\pi^0}  &=& \Frac{\alpha}{m_\pi} \, \Frac{1}{(4\pi F)^4} \,
\bigg\{ \,
\left(  \cO(p^6) \mbox{ loops} \right) \, +
\,\cO(m_\pi^2)\bigg\}\, ,
\eear
where the $\cO(m_\pi^4)$ and $\cO(m_\pi^2)$ terms at the end of each equation represent
the contributions of $\cO(p^8)$ and higher in $\chi$PT.

For the $\gamma\gamma\to\pi^+\pi^-$ polarizabilities one has the chiral
expansion~\cite{Burgi:1996,Gasser:2006}
\bear
(\alpha_1-\beta_1)_{\pi^+} &=& \Frac{\alpha}{m_\pi}\,\Frac{1}{(4\pi F)^2} \,
\bigg\{  \,   \Frac{2 \bar{\ell}_\Delta}{3}\,\,\, +\,\, \, \left( a_1^r+ 8 b^r + \cO(p^6) \mbox{ loops}\right) \,
 \Frac{m_\pi^2}{(4\pi F)^2} \, \, + \,\,\cO(m_\pi^4)\bigg\}\, ,
\nn\\
(\alpha_1+\beta_1)_{\pi^+} &=& \Frac{\alpha}{m_\pi}\, \Frac{m_\pi^2}{(4\pi F)^4} \,
\bigg\{ \, \left(8b^r +  \cO(p^6) \mbox{ loops} \right)
\,  \, + \,\, \cO(m_\pi^2)\bigg\}\, ,
\nn\\
(\alpha_2-\beta_2)_{\pi^+} &=& \Frac{\alpha}{m_\pi} \,
 \Frac{1}{m_\pi^2 (4\pi F)^2}\,\bigg\{ \,2\, +\,
\left(12 a_2^r -24 b^r +  \cO(p^6) \mbox{ loops} \right)\, \Frac{m_\pi^2}{(4\pi F)^2}  \, +
\,\cO(m_\pi^2)\bigg\}\, ,
\nn\\
(\alpha_2+\beta_2)_{\pi^+}  &=& \Frac{\alpha}{m_\pi} \, \Frac{1}{(4\pi F)^4} \,
\bigg\{ \,
\left(   \cO(p^6) \mbox{ loops} \right) \, +
\,\cO(m_\pi^2)\bigg\}\, ,
\eear
with $\bar{\ell}_\Delta  =  192 \, \pi^2 (L_9(\mu)+L_{10}(\mu))$.
Again, the $\cO(m_\pi^4)$ and $\cO(m_\pi^2)$ terms
at the end of each equation
represent the contributions of
$\cO(p^8)$ and higher in $\chi$PT.
Since the $a_1$ parameter is out of the  reach  in our massless quark holographic approach,
we  focus  on  the combinations
$(\alpha_1+\beta_1)$ and $(\alpha_2\pm \beta_2)$.

At large $N_C$, in our holographic approach based on  the $m_\pi\to 0$ limit,  we have:
\bear
H_{++}(s,t=m_\pi^2) &=& \Frac{a_2}{(4\pi F)^4}\bigg[ \Frac{2s}{3}\, +\, \cO(s^2)\bigg]\, ,
\nn\\
H_{+-}(s,t=m_\pi^2) &=& \Frac{b \, m_\pi^2}{(4\pi F)^4}
\bigg[ 8 \, -\, \Frac{96 L_1 s}{F^2}\, +\, \cO(s^2)\bigg]\, .
\eear
 This  leads to the large $N_C$ determinations of the polarizabilities,
\bear
(\alpha_1+\beta_1) &=& \Frac{\alpha}{m_\pi}\, \Frac{8 b \,m_\pi^2}{(4\pi F)^4}
\,  \, + \,\, \cO(m_\pi^3)\, ,
\nn\\
(\alpha_2-\beta_2) &=& \Frac{\alpha}{m_\pi} \,  \Frac{8 a_2}{  (4\pi F)^4}  \, +
\,\cO(m_\pi) \, ,
\nn\\
(\alpha_2+\beta_2)  &=& \,-\, \Frac{\alpha}{m_\pi} \,
\Frac{1152 L_1  b}{F^2}\,  \Frac{ m_\pi^2 }{(4\pi F)^4} \,  + \,\cO(m_\pi^3)\, ,
\eear
where we used  the holographic prediction $a_2 =6 \, b$.
Both the neutral and charged channels have the same structure and one must use the
corresponding $a_2$ and $b$ parameters.
Notice that  in the  large $N_C$ limit the polarizability  $(\alpha_2+\beta_2)$ starts at $\cO(p^8)$.
In real world, the leading tree-level $\cO(p^8)$ contribution  obtained here competes  with
the $1/N_C$ suppressed $\cO(p^6)$ loops.

It is convenient to keep track of the different contributions, and observe
at which chiral order each of the polarizabilities  begins.
The first non-vanishing contribution appears at $\cO(p^4)$
and  comes  from one-loop diagrams.
Indeed, only $(\alpha_2-\beta_2)$ is different from zero at this order,  and the other polarizabilities
$(\alpha_1+\beta_1)$ and $\alpha_2+\beta_2)$
vanish~\cite{Bijnens:1988,Burgi:1996,Gasser:2005,Gasser:2006}.
The chiral expansion for $(\alpha_1+\beta_1)$
begins at $\cO(p^6)$
(loop+tree-level)~\cite{Burgi:1996,Gasser:2005,Gasser:2006}.
The polarizability $(\alpha_2+\beta_2)$ also starts
at this order but only via
loops, since the  first tree-level contribution appears at $\cO(p^8)$~\cite{Burgi:1996,Gasser:2005,Gasser:2006}.

At large $N_C$ the first contribution to
both $(\alpha_1 +  \beta_1)_{\pi^0}$ and $(\alpha_2 -\beta_2)_{\pi^0}$ starts at $\cO(p^6)$.
The polarizability   $(\alpha_2 + \beta_2)_{\pi^0}$  is even more suppressed at large $N_C$,
starting at $\cO(p^8)$.
In Table~\ref{tab.pol-comparison1} we  see how the values of the polarizabilities
evolve as we include higher  chiral orders.
In the first column  we provide the one-loop $\cO(p^4)$
contributions~\cite{Bijnens:1988}
(the $\cO(p^2)$ pion Born term
is explicitly removed in the charged channel definitions and absent in the neutral one).  Then
we add the lowest order contribution from tree-level resonance exchanges from our holographic
Lagrangian, $\cO(p^6)$ for $(\alpha_1+\beta_1)$
and $(\alpha_2-\beta_2)$ and $\cO(p^8)$ in the case of $(\alpha_2+\beta_2)$.
Finally, in the last column we also sum up the $\cO(p^6)$ loop contribution.
We provide three numbers: the first one is given by $\bar\ell_3$
and $\bar\ell_4$ from~\cite{Gasser:2005,Gasser:2006} and the values of $\bar\ell_{1,2,\Delta}$
extracted from  $L_{1,2,3,9,10}(\mu)$
estimated from the ``Cosh'' model at $\mu=770$~MeV; the second one is similar but with
$\bar\ell_{1,2,\Delta}$ estimated from the Hard-Wall model; for the third number (in brackets)
we have used the values of the $\cO(p^4)$ LEC's  from~\cite{Gasser:2005,Gasser:2006}
in the $\cO(p^6)$ loop contribution.
In the neutral channel one can see that the resonance contributions seem to be slightly
dominant with respect to the $\cO(p^6)$ loops.  However, vector resonance exchanges in
the $\gamma\gamma\to\pi^+\pi^-$ amplitude carry a $\frac{1}{9}$  suppression factor and
we found them  of the same numerical size as the $\cO(p^6)$ loops.

In Table~\ref{tab.pol-comparison2} we compare the  result
in the last column of Table~\ref{tab.pol-comparison1}
to other determinations.
In particular, we quote the outcome
of dispersive analyses such as the Muskhelishvili-Omn\`es (MO) relation
in terms of the $\pi\pi$--scattering phase-shifts~\cite{Moussallam:2010}
and the Roy-Steiner equations~\cite{Hoferichter:2011wk},
together with the result of the $\chi$PT computation
at $\cO(p^6)$~\cite{Gasser:2005,Gasser:2006}.
We obtain an overall agreement between the holographic determination
and the dispersive  and chiral computations.
Further comparisons can be carried out with previous experimental and theoretical results~\cite{Pennington-rev}
collected in Table~\ref{tab.pol-comparison3}

We  remark again that in our computation we have considered the charge matrix
Q=diag$(\frac{2}{3},-\frac{1}{3} ,-\frac{1}{3})$  in the calculation of our large--$N_C$
estimates of the LEC's. Nonetheless, we have found a relatively good numerical agreement
with the next-to-next-to-leading order (NNLO) $\chi$PT calculations from Refs.~\cite{Gasser:2005,Gasser:2006},
which rather considered the $SU(2)$ charge matrix Q=diag$(\frac{1}{2},-\frac{1}{2} )$.

\begin{table}[!h]
\centering
\begin{tabular}{|c|c|c|c|}
\hline
& $\cO(p^4)$
&    $\cO(p^4)$
&    $\cO(p^4)$
\\
&
& + resonance
& + reson. (hologr.)
\\
&
&  (hologr.)
& + $\cO(p^6)$ loops
 \\ \hline
$(\alpha_1+\beta_1)_{\pi^0}$
& 0   & 0.58 & ~0.75~~;~~0.74~~;~~[0.69]~
\\
\hline
$(\alpha_2-\beta_2)_{\pi^0}$
& 20.73  & 27.67 &   ~30.18~~;~~29.98~~;~~[34.65]~
\\
\hline
$(\alpha_2+\beta_2)_{\pi^0} $
& 0 & -0.24 &  ~-0.16~~;~~-0.17~~;~~[-0.20]~
 \\ \hline
&&&
 \\ \hline
$(\alpha_1+\beta_1)_{\pi^+}$
& 0 &  0.06 &  ~0.16~~;~~0.16~~;~~[0.08]~
\\
\hline
$(\alpha_2-\beta_2)_{\pi^+}$
& 11.96 &  12.65 & ~14.23~~;~~14.12~~;~~[17.08]~
\\
\hline
$(\alpha_2+\beta_2)_{\pi^+} $
& 0 & -0.02  & ~0.03~~;~~0.02~~;~~[-0.02]~
\\   \hline
\end{tabular}
\caption{{\small
Polarizabilities in units of $10^{-4}$~fm$^3$ for $\alpha_1$, $\beta_1$,
and $10^{-4}$~fm$^5$ for $\alpha_2$, $\beta_2$. They are provided at $\cO(p^4)$ in the second column whereas
in the third column we add the resonance contribution from the holographic models, which begins at $\cO(p^6)$ for $(\alpha_1+\beta_1)$
and $(\alpha_2-\beta_2)$, and at $\cO(p^8)$ in the $\chi$PT expansion for $(\alpha_2-\beta_2)$.
The $\cO(p^6)$ loop contributions are finally added in the last column.
}}
 \label{tab.pol-comparison1}
 \end{table}

\begin{table}[!h]
\centering
\begin{tabular}{|c|c|c|c|}
\hline
&    $\cO(p^4)$
&
Dispersive
& NNLO
\\
& + reson.  (hologr.)
& analysis
&$\chi$PT
\\
&  + $\cO(p^6)$ loops
&   \cite{Moussallam:2010,Hoferichter:2011wk}
&\cite{Gasser:2005,Gasser:2006}
 \\ \hline
$(\alpha_1+\beta_1)_{\pi^0}$
& ~0.75~~;~~0.74~~;~~[0.69]~
& $1.22\pm 0.12 \pm 0.03$
& $1.1\pm 0.3$
\\
\hline
$(\alpha_2-\beta_2)_{\pi^0}$
&     ~30.18~~;~~29.98~~;~~[34.65]~
& $32.1\pm 0.9\pm 1.9$
& $37.6\pm 3.3$
\\
\hline
$(\alpha_2+\beta_2)_{\pi^0} $
&  ~-0.16~~;~~-0.17~~;~~[-0.20]~
&  $-0.19\pm 0.02\pm 0.01$
& 0.04
 \\ \hline
 &&&
 \\ \hline
$(\alpha_1+\beta_1)_{\pi^+}$
&  ~0.16~~;~~0.16~~;~~[0.08]~
&  $0.19\pm 0.09\pm 0.03$
& 0.16 \, [0.16]
\\
\hline
$(\alpha_2-\beta_2)_{\pi^+}$
&  ~14.23~~;~~14.12~~;~~[17.08]~
&  $14.7\pm 1.5\pm 1.4$
& 16.2\, [21.6]
\\
& &
$15.3 \pm 3.7$
&
\\
\hline
$(\alpha_2+\beta_2)_{\pi^+} $
&   ~0.03~~;~~0.02~~;~~[-0.02]~
& $0.11\pm 0.03\pm 0.01$
& -0.001
\\   \hline
\end{tabular}
\caption{{\small
Our holographic predictions for the polarizabilities,  provided
in the second column
(same outcomes as in the last column in Table~\ref{tab.pol-comparison1}),
are compared to
the dispersive results from Roy-Steiner equations~\cite{Hoferichter:2011wk}
(second value for $(\alpha_2-\beta_2)_{\pi^+}$ in the third column)
and the MO representation~\cite{Moussallam:2010} (all the remaining outcomes in the
third column).
They are also compared to the NNLO $\chi PT$
analyses~\cite{Gasser:2005,Gasser:2006}, provided in the last column.
Units are the same as in Table~\ref{tab.pol-comparison1}.
}}
 \label{tab.pol-comparison2}
 \end{table}

\begin{table}[!h]
\centering
\begin{tabular}{|c|c|c|c|c|c|}          
\hline
 &   CELLO  &  MARK-II        
& Crystal    & Vector & Sum
\\
& &        
& Ball  & exchanges   & rules
\\
&  \cite{Kaloshin:1994,CELLO}  &  \cite{Kaloshin:1994,MARK-II}          
& \cite{Kaloshin:1994,CrystalBall:1990} & \cite{Ko:1990,Babusci:1993}&~\cite{Filkov:2005}
 \\
\hline
$(\alpha_1+\beta_1)_{\pi^0}$
 &  &              
& $1.00\pm 0.05$    & 0.83 &   $0.802\pm 0.035$
\\
\hline
$(\alpha_2-\beta_2)_{\pi^0}$
&    & &             
& &  $39.72\pm 8.01$
\\
\hline
$(\alpha_2+\beta_2)_{\pi^0} $
&    & &                
& &  $-0.171\pm 0.067$
 \\ \hline
  &&&               
  & &
\\
\hline
$(\alpha_1+\beta_1)_{\pi^+}$
&  $0.30\pm 0.04$  &   $0.22\pm 0.06$                  
&   &   0.07 & $0.166\pm 0.024$
\\
\hline
$(\alpha_2-\beta_2)_{\pi^+}$
&    & &                     
& &  $25.75\pm 7.03$
\\
\hline
$(\alpha_2+\beta_2)_{\pi^+} $
&    &  &              
& & $0.121\pm 0.064$
\\   \hline
\end{tabular}
\caption{{\small
Pion polarizabilities from experimental measurements~\cite{Kaloshin:1994,CELLO,MARK-II,CrystalBall:1990} and
theoretical analyses~\cite{Ko:1990,Babusci:1993,Filkov:2005}.
Units are the same as in Table~\ref{tab.pol-comparison1}.
}}
 \label{tab.pol-comparison3}
 \end{table}
\section{Conclusions}

Following previous analyses~\cite{Son:2010vc,Colangelo:2012ip}, we have determined a novel set of relations
between QCD matrix elements using holographic models where chiral symmetry is broken through IR b.c.'s.
We have focused  on the scattering amplitudes of pions and photons,
finding that the three processes  $\pi\pi\to\pi\pi$, $\gamma\gamma\to\pi\pi$ and $\gamma\to\pi\pi\pi$
involve a single scalar function $h(Q^2)$.  This function is given by a suitable 5D integral of
the EoM Green's function and accepts the usual decomposition in terms of resonance exchanges. Furthermore,
in the considered  processes
only the vector mesons  contribute (scalars and resonances of spin $S\geq 2$ are not included in the present approach).

In a  detailed phenomenological analysis of   $\gamma\gamma\to\pi\pi$ we have
 found an overall agreement with the experimental cross section for a broad range of energy. Likewise,
the computed  polarizabilities at low energies show
a fair agreement between the holographic approach,  previous computations and experiment.

\section*{Acknowledgments}
We thank
Fulvia De Fazio, Juerg Gasser, Floriana Giannuzzi, Mihail Ivanov and Stefano Nicotri
for  useful discussions.
This work is partially supported by the Italian Miur PRIN 2009,
the Universidad CEU Cardenal Herrera grant PRCEUUCH35/11,  the MICINN-INFN fund AIC-D-2011-0818,
and by the National Natural Science Foundation of China under Grant No. 11135011.

\appendix
\section{Constraints on the background functions}\label{app.constraints}
Here we provide  constraints on the background functions $f^2(z)$ and $g^2(z)$. As proposed in ref.~\cite{Son:2003et},
these functions must be invariant under the reflection of $z$
in order to properly define the parity. More  constraints come from the results for the low-energy constants.

With the resonance decomposition of the gauge potential~(\ref{eq.Amu-decomposition1}), up to $\cO(p^4)$ the 5D Yang-Mills action reduces to the $\chi PT$  Lagrangian~\cite{Hirn:2005nr,Sakai:2004cn,Colangelo:2012ip}:
\begin{eqnarray}
S_2[\pi]+S_4[\pi]&=& \Int d^4x\, \bigg[
\frac{F^2}{4}<u_\mu u^\mu>\nonumber\\
        &&\qquad\qquad +L_1 <u_\mu u^\mu>^2+L_2 <u_\mu u_\nu><u^\mu u^\nu> +L_3 <u_\mu u^\mu u_\nu u_\nu>\nonumber\\
        &&\qquad\qquad  -iL_9<f_{+\mu\nu}u^\mu u^\nu>
+\frac{L_{10}}{4}<f_{+\mu\nu}f_+^{\mu\nu}-f_{-\mu\nu}f_-^{\mu\nu}>
\nn\\
        &&\qquad\qquad +\frac{H_1}{2}<f_{+\mu\nu}f_+^{\mu\nu}+f_{-\mu\nu}f_-^{\mu\nu}>\,
        \bigg]\, .\label{app:a1}
\end{eqnarray}
The  low-energy constants  in (\ref{app:a1}) are given by the 5D integrals
\begin{eqnarray}
F^2&=&4\left(\int_{-z_0}^{z_0}\frac{\mathd z}{f^2(z)}\right)^{-1} \, ,
\nn \\
L_1&=&\frac{1}{2} L_2=-\frac{1}{6}L_3=\frac{1}{32}\int_{-z_0}^{z_0} \frac{(1-\psi_0^2)^2}{g^2(z)}  \, \mathd z \, ,
\nn\\
L_9&=&-L_{10}=\frac{1}{4}\int_{-z_0}^{z_0} \frac{1-\psi_0^2}{g^2(z)}  \, \mathd z \, ,
\\
H_1&=&-\frac{1}{8}\int_{-z_0}^{z_0} \frac{1+\psi_0^2}{g^2(z)}   \, \mathd z \, .
\nn
\end{eqnarray}
We demand that all these integrals except $H_1$ are finite. For $H_1$, which is the coefficient of the kinetic term of the external sources,
we require it to be  divergent.  From the finiteness of the pion decay constant $F$ we find the  solution
\begin{equation}
\psi_0(z)=\frac{F^2}{2}\int_0^z \frac{1}{f^2(z)} \mathd z \,\,\, ,
\end{equation}
which satisfies the equation of motion with  boundary conditions $\psi_0(\pm z_0)=\pm~1$. Since $\psi_0(z)$ is
a monotonic function of $z$, we can choose it as the coordinate parameter through a coordinate transformation in $z$. Defining $y=\psi_0(z)$, it is not difficult to find the new background functions
\begin{equation}
{\tilde f}^2(y)=\frac{F^2}{2},\,\,\,\,{\tilde g}^2(y)=\frac{F^2}{2} \frac{g^2(z(y))}{f^2(z(y))},
\end{equation}
together with the boundaries $y=\pm 1$. It turns out that this coordinate system is convenient in many respects, both for theoretical derivations and numerical calculations.
In this coordinate system the integral for  $F$ becomes trivial, and the other integrals can be expressed as
\begin{eqnarray}
L_1&=&\frac{1}{32 }\int_{-1}^{1} \frac{(1-y^2)^2}{{\tilde g}^2(y)}  \, \mathd y \,\, ,
\nn\\
L_9&=&\frac{1}{4}\int_{-1}^{1} \frac{1-y^2}{{\tilde g}^2(y)}  \, \mathd y \,\, ,
\\
H_1&=&-\frac{1}{8}\int_{-1}^{1} \frac{1+y^2}{{\tilde g}^2(y)}   \, \mathd y \,\, .
\nn
\end{eqnarray}
Requiring that $L_1$ and $L_9$ are finite and $H_1$ divergent, we get the constraint near the boundaries
\begin{equation}
(1-y^2)^2<{\tilde g}^2(y)\leq C(1-y^2) \label{eq.fg-constraint}
\end{equation}
with $C$ a constant. Actually, the explicit boundary behavior of the function, ${\tilde g}^2(y)$, can be used to clarify the ultraviolet property of different models. Among the models shown in the next appendix, this function behaves as $(1-y^2)^0$ in the flat model, related to a convergent value of $H_1$, while in the Sakai-Sugimoto model, it goes as $(1-y^2)^{4/3}$. As for all the asymptotic anti-de Sitter backgrounds, the equality in the above relation is exactly satisfied, e.g., in the ``cosh'' and Hard-Wall models.

In the new coordinate system, the quantity $\mH$ defined in Eq.~(\ref{eq.H}) simplifies as
\begin{equation}
\mH=\frac{1}{4}\int_{-1}^{1}~{\tilde g}^2(y) \,\mathd y \,\, ,
\end{equation}
and, with the constraint (\ref{eq.fg-constraint}), it is finite.

\section{Holographic models} \label{app.holographic-models}

We have used four different holographic models, defined by the functions $f^2(z)$ and $g^2(z)$ and by the value of $z_0$.
Here we list their details in each model. The expressions of the wave functions solutions of the equation of motion,
and  of other   quantities like $F$, the   couplings  and the mass spectrum, can be found in the appendix of ref.~\cite{Colangelo:2012ip}.

{``}Flat'' background~\cite{Son:2003et}:
\begin{equation}
f^2(z)=\frac{\Lambda^2 }{g_5^2} \,\,\, , ~~g^2(z)=g_5^2\,\,\,\ , ~~z_0=1.
\end{equation}

{``}Cosh'' model ~\cite{Son:2003et}:
\begin{equation}
f^2(z)=\frac{\Lambda^2 \cosh^2(z)}{g_5^2}\,\,\, , ~~g^2(z)=g_5^2\,\,\, ,~~ z_0=\infty.
\end{equation}

{``}Hard-wall" model~\cite{Hirn:2005nr}:
\begin{equation}
f^2(z)= \frac{1}{g_5^2(z_0-|z|)}\,\,\, ,~~ g^2(z)=g_5^2 (z_0-|z|)\,\,\, , ~~z_0<\infty.
\end{equation}

{``}Sakai-Sugimoto" model~\cite{Sakai:2004cn,Sakai:2005yt}:
\begin{equation}
f^2(z)=\frac{\Lambda^2 (1+z^2)}{g_5^2}\,\,\, , ~~g^2(z)=g_5^2 (1+z^2)^{1/3}\,\,\, , ~~z_0=\infty.
\end{equation}

\section{Isospin and partial-wave projection in $\pi\pi$--scattering}

The amplitude $A(s,t,u)$ provides the scattering amplitudes $T^I$ of modes with definite
isospin $I=0,1,2$~\cite{op4-chpt,op6-pipi-scat}:

\bear
T^0(s,t,u) &=&  3 A(s,t,u) + A(t,s,u) + A(u,s,t)\, ,
\nn\\
T^1(s,t,u) &=& A(t,s,u) -A(u,s,t)\, ,
\nn\\
T^2(s,t,u) &=& A(t,s,u) + A(u,s,t)\, .
\eear

In the isospin $I=2$ amplitude, there are no resonances in the s--channel, only exchanges in the crossed channels.
This simplifications makes the amplitude particularly interesting for the study of its partial waves, which in the
massless quark limit have the form~\cite{op4-chpt,op6-pipi-scat}

\bear
T^I_J(s) &=& \Frac{1}{32\pi s}\Int_{-s}^0 {\rm dt}\,\, P_J\bigg(1+\frac{2t}{s}\bigg) \, T^I(s,t,u)
\nn\\
&&   \,\,\,=\,\,\,
\Frac{1}{64\pi}\Int_{-1}^1 {\rm dx}\,\, P_J(x) \, T^I(s,-s (1-x)/2, -s (1+x)/2)
\nn\\
&&   \,\,\,=\,\,\,
\Frac{1}{32\pi}\Int_0^1 {\rm dy}\,\, P_J(1-2y) \, T^I(s,-s y, -s (1-y))\, ,
\nn\\
\eear
with $u=-s-t$ (in the chiral limit), $y=(1-x)/2= -t/s$   and $P_J$ the Legendre polynomials.

\section{Expressions for the relevant $\cO(p^6)$ LEC's}\label{app.LECe}

Here we provide the expressions of some $\cO(p^6)$ LEC's derived
in ref.~\cite{Colangelo:2012ip}, which have been used in the calculation of the polarizabilities. They are summarized in Table~\ref{tab:LECe},
where the following definitions have been used:
\begin{equation}
 S_{VV}=\sum_{n=1}^\infty\frac{a_{Vv^n}^2}{m_{v^n}^2}\, ,\qquad
 S_{V\pi\pi}=\sum_{n=1}^\infty\frac{a_{Vv^n}b_{v^n\pi\pi}}{m_{v^n}^2}\, ,\qquad
 S_{AA}=\sum_{n=1}^\infty\frac{a_{Aa^n}^2}{m_{a^n}^2}\,.
 \label{eq:sum}
 \end{equation}
The coupling $a_{Aa^n}$ comes from the interaction
\begin{equation}
S_{\rm{YM}}\bigg|_{{\rm 1-res.}}\supset \qquad  \,   \bra  \frac{f^{\mu\nu}_-}{2}
 (\nabla_\mu a_\nu^n-\nabla_\nu a_\mu^n)a_{Aa^n} \ket,
\end{equation}
with $f_{-}^{\alpha\beta}=\xi_L^\dagger F_L^{\alpha\beta} \xi_L -
\xi_R^\dagger F_R^{\alpha\beta} \xi_R$.
The explicit contributions given in terms of $F$ and $N_C$
come from the diagrams with two parity-odd vertices,
while the other terms come from diagrams with two even-parity
vertices. From the expressions in
Tab. \ref{tab:LECe} one easily recovers the results
(\ref{eq.neutral-a2-b-holography}),~(\ref{eq.neutral-a2-b-holography2}),
(\ref{eq.charged-a2-b-holography}) and~(\ref{eq.charged-a2-b-holography2}).

\begin{table}[t!]
\centering
\begin{tabular}{|c|c|}
\hline
     ~~~~~$C_{53}$~~~~~  & ~~~~~~$\frac{3}{16}S_{AA}-\frac{1}{16}S_{V\pi\pi}-\frac{3}{16}S_{VV}+\frac{N_C^2}{3072\pi^4 F^2}$ ~~~~~~ \\\hline
   $C_{55}$ &  $-\frac{3}{16}S_{AA}+\frac{1}{16}S_{V\pi\pi}+\frac{3}{16}S_{VV}+\frac{N_C^2}{3072\pi^4 F^2}$\\\hline
   $C_{56}$  & $-\frac{3}{8}S_{AA}-\frac{1}{8}S_{V\pi\pi}+\frac{3}{8}S_{VV}-\frac{N_C^2}{1536\pi^4 F^2}$   \\\hline
   $C_{57}$ &  $-\frac{1}{8}S_{AA}+\frac{1}{4}S_{V\pi\pi}+\frac{1}{8}S_{VV}$   \\\hline
   $C_{59}$  & $\frac{1}{4}S_{AA}-\frac{1}{16}S_{V\pi\pi}-\frac{1}{4}S_{VV}-\frac{N_C^2}{3072\pi^4 F^2}$ \\\hline
        $C_{78}$ &  $-\frac{1}{4}S_{AA}+\frac{1}{16}S_{V\pi\pi}+\frac{1}{4}S_{VV}$ \\\hline
           $C_{87}$ &  $-\frac{1}{8}S_{AA}+\frac{1}{8}S_{VV}$ \\\hline
            $C_{88}$  & $-\frac{1}{8}S_{V\pi\pi}$ \\\hline
\end{tabular}
\caption{Holographic predictions for some of the $\cO(p^6)$ LEC's
in the even sector~\cite{Colangelo:2012ip}.}
\label{tab:LECe}
\end{table}

%

\section{Contribution to $\gamma\gamma\to \pi\pi$ from $\cO(p^4)$ diagrams }
\label{app.Op4}

In the neutral channel $\gamma\gamma\to\pi^0\pi^0$, the only $\cO(p^4)$ contribution
appears at the one loop level~\cite{Bijnens:1988,Gasser:2005}:
\bear
\Delta A(s,t,u)^{\gamma\gamma\to\pi^0\pi^0}_{\rm \cO(p^4)}
&=& \Frac{4\bar{G}_\pi(s)}{F_\pi^2} \, \bigg(1-\Frac{m_\pi^2}{s}\bigg)\, ,
\nn\\
\Delta B(s,t,u)^{\gamma\gamma\to\pi^0\pi^0}_{\rm \cO(p^4)}  &=&  0\, ,
\eear
with
\bear
\bar{G}_\pi(s) &=& -\,\Frac{1}{16\pi^2}\,\bigg[ \,1\, +\, \Frac{m_\pi^2}{s}
\, \bigg(\ln\Frac{\rho_\pi+1}{\rho_\pi-1}\bigg)^2\, \bigg]\, ,
\eear
given by the phase-space factor $\rho_\pi=\sqrt{1-4m_\pi^2/s}$.

In the $\gamma\gamma\to \pi^+\pi^-$ channels the $\cO(p^4)$ diagrams
contribute both at one loop and tree-level~\cite{Bijnens:1988,Gasser:2006}:
\bear
\Delta A(s,t,u)^{\gamma\gamma\to\pi^+\pi^-}_{\rm \cO(p^4)}
&=& \Frac{2\bar{G}_\pi(s)}{F_\pi^2} \,\,\,+\,\,\,
\Frac{\bar\ell_\Delta}{24\pi^2 F_\pi^2}\, ,
\nn\\
\Delta B(s,t,u)^{\gamma\gamma\to\pi^+\pi^-}_{\rm \cO(p^4)}  &=&  0\, ,
\eear
with the tree-level term $\bar\ell_\Delta=\bar{\ell}_6-\bar\ell_5
= 192\pi^2 (\ell_5(\mu)-\ell_6(\mu)/2)$. However,
in the type of holographic models studied in this work
one always has $\ell_5 -\ell_6/2 = L_{10}+L_9 =0$
at large  $N_C$~\cite{Hirn:2005nr,Sakai:2005yt}.

%

%
%
%

\end{document}